\documentclass[aps,prl,twocolumn,showpacs,preprintnumbers,amsmath,amssym,superscriptaddress]{revtex4}
\usepackage{graphicx} \usepackage{bm,ulem,color,textcomp}
\usepackage{amsmath,amsfonts,amssymb} 
 \newcommand{\ignore}[1]{}

\begin{document} 
\title{Temporal Griffiths Phases}
\author{Federico Vazquez} \affiliation{Max-Planck-Institut f\"ur
  Physik komplexer Systeme, N\"othnitzer Str. 38, 01187 Dresden,
  Germany} \author{Juan A. Bonachela} \affiliation{Dept. of Ecology
  and Evolutionary Biology, Princeton Univ., Princeton, NJ 08544-1003,
  USA} \author{Crist\'obal L\'opez} \affiliation{IFISC, Inst. de
  F{\'\i}sica Interdisciplinar y Sistemas Complejos (CSIC-UIB),
  E-07122 Palma de Mallorca, Spain.}  \author{Miguel A. Mu\~noz}
\affiliation{
Inst. Carlos I de F{\'\i}sica Te{\'o}rica y Computacional, Universidad
de Granada, 18071 Granada, Spain}
\date{\today}

\begin{abstract}
  Disorder is an unavoidable ingredient of real systems.
  Spatial disorder generates {\it Griffiths phases} (GPs) which, in
  analogy to critical points, are characterized by a slow relaxation
  of the order parameter and divergences of quantities such as the
  susceptibility. However, these singularities appear in an
  extended region of the parameter space and not just at a (critical)
  point, i.e. there is {\it generic scale invariance}.  Here, we study
  the effects of temporal disorder, focusing on systems with absorbing
  states.  We show that for dimensions $d \geq 2$ there are
  {\it Temporal Griffiths phases (TGPs)} characterized by generic
  power-law spatial scaling and generic divergences of the
  susceptibility. TGPs turn out to be a counterpart of GPs, but with space and time playing reversed roles. TGPs
  constitute a unifying concept, shedding light on the non-trivial
  effects of temporal disorder.
 \end{abstract}

 \maketitle 

 Quenched disorder affects the behavior of particle systems, altering
 critical properties and introducing new universality
 classes. Non-magnetic impurities in magnetic systems or defects in
 type-I superconductors are typical examples of this
 \cite{Geoff}. Moreover, novel phases, with phenomenology unheard-of
 in pure systems, can be induced by spatial disorder. This is the case
 of {\it Griffiths phases} (GPs) appearing in classical, quantum, and
 non-equilibrium disordered systems \cite{Griffiths,GP,Vojta}. GPs,
 which are of relevance in condensed matter physics as well as in
 other contexts \cite{Vojta,Odor}, are regions of the phase space
 \texttwelveudash actually a sub-region of the disordered phase
 \texttwelveudash characterized by extended singularities of the
 thermodynamic potentials and, as a consequence, generic divergences
 of magnitudes such as the susceptibility \cite{Griffiths}.
 Furthermore, GPs are characterized by an anomalously slow (power-law
 or stretched exponential) relaxation to zero of the order parameter
 (and of other time-dependent quantities) which contrasts with the
 fast (exponential) decay typical of pure systems. Such an anomalous
 relaxation in the disordered phase occurs owing to the presence of
 rare regions where the disorder is such that the system is locally in
 its ordered phase and, hence, a potential barrier has to be overcome
 for it to relax. The convolution of different, exponentially rare,
 sizes with exponentially large decaying times gives rise to an
 overall slowing down of the system's dynamics, which typically
 becomes algebraic in time \texttwelveudash with continuously varying
 exponents \texttwelveudash and logarithmic at the critical point (see
 below for more details) \cite{GP,Vojta}.  Divergences in the
 potentials together with slow relaxation are two features strongly
 reminiscent of criticality and its concomitant
 scale-invariance. However, in GPs these traits appear not just at a
 critical point but in a broad extended region, providing a robust
 mechanism to justify some cases of scale-invariance in Nature
 \cite{SI}.

 The modeling of some problems in physics, chemistry or ecology
 requires parameters to be disordered in {\it time} rather than in
 {\it space} \cite{foot}. This is the case of magnetic systems under a
 fluctuating external field \cite{Geoff}, or of ecological populations
 under changing environmental conditions \cite{Leigh}.  In general,
 temporal fluctuations in the parameters can shift critical points
 \cite{Hors} and affect universal features both in equilibrium
 \cite{Juanjo} and non-equilibrium systems \cite{Jensen}. In a
 pioneering work, Leigh showed that, in one-variable (mean-field)
 models of stochastic populations, environmental noise changes the
 system mean lifetime (time to reach the absorbing state) from
 exponential to a power-law in system-size \cite{Leigh,Kamenev}.  This
 result inspired us to systematically explore the role of temporal
 disorder in spatially extended systems (beyond mean-field) and to
 study if rare temporal regions induce new phases analogous to
 spatial-disorder induced GPs.  Do {\it temporal Griffiths phases\;}
 exist? If so, which properties do they have?  Do they exhibit any
 type of generic scale invariance?

\begin{figure}[t]
 \includegraphics[width=0.32\textwidth]{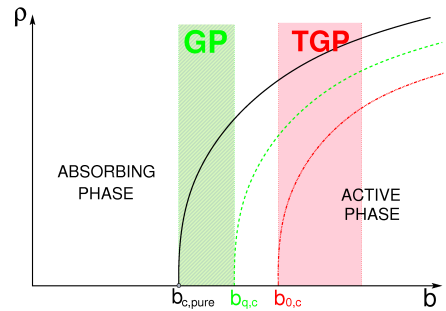}
 \caption{Schematic phase diagram for the pure contact process (CP)
   (solid line) the CP with quenched disorder (dashed line), and the
   CP with temporal disorder (dot-dashed line). For the second, a
   Griffiths phase appears within the absorbing region, while for the
   third a Temporal Griffiths phase appears (for $d>1$) within the
   active region.  The actual locations of the critical points may
   depend on noise intensity and dimension.}
\label{Phases}
\end{figure}

 To tackle these questions, we start by analyzing a specific model
 with absorbing states: the contact process (CP) \cite{AS}, in the
 presence of temporal disorder. In the CP, each site of a
 $d$-dimensional lattice can be either occupied $z({\bf x})=1$
 (active) or vacant $z({\bf x})=0$. At each time step, an active site
 is randomly chosen and, with probability $b$, it converts into active
 a nearest neighboring site (provided it was empty), while with
 probability $1-b$ it is declared empty.  Time, $t$, is then increased
 by $1/N(t)$, where $N(t)$ is the total number of active sites.  The
 ``pure'' CP is critical only at some dimension-dependent value
 $b_{c,pure}(d)$ separating an active from an absorbing phase (see
 \cite{AS} and the schematic diagram in Fig.\ref{Phases}). This phase
 transition, occurring at $b_{c,pure} \approx 0.767$, $0.622$, and
 $0.5$ for $d=1$, $d=2$ and $d=\infty$, respectively, lies in the very
 robust {\it directed percolation} universality class \cite{AS}.  For
 the spatially-disordered case, b is replaced by $b({\bf
   x})$; in this case, a GP emerges between $b_{c,pure}$ and the
 critical point of this quenched version, $b_{q,c}>b_{c,pure}$
 (see Fig. \ref{Phases}).

 Temporal disorder is implemented by allowing $b$ to be a
 time-dependent random variable, $b \rightarrow b(t)$, for all ${\bf
   x}$. In the simplest (uncorrelated) case, $b$ takes a random value
 extracted at each Monte Carlo step (i.e. whenever the integer value
 of $t$ increases) from some distribution of mean $b_0$ and width
 $\sigma$. Correlated fluctuations can also be implemented, by
 allowing $b$ to obey an Ornstein-Uhlenbeck dynamics
 \cite{Gardiner,Kamenev}. This temporally-disordered contact process
 ({\bf TD-CP}) is similar to the Jensen's model in
 \cite{Jensen}. Following the instantaneous value of $b$, the system
 shifts between the tendencies to be active ($b(t) >b_{c,pure}$) or
 absorbing ($b(t) <b_{c,pure}$), provided that the disorder
 distribution is broad enough (see Fig.\ref{Phases} for a schematic
 diagram). Owing to fluctuations, any finite system is, however,
 condemned to end up in the absorbing state, either for fixed and
 changing $b$ \cite{Gardiner}.
However, the mean lifetime $\tau(N)$ grows exponentially with system size $N$
 (Arrhenius law \cite{Gardiner})  in the pure system case,
 making it stable in the thermodynamic
 limit. Instead, in the TD-CP, as $b(t)$ can be adverse
 ($b(t)<b_{c,pure}$) for arbitrarily long time intervals, $\tau(N)$ is
 expected to be significantly reduced. But, does $\tau(N)$ still
 diverge for $N \rightarrow \infty$?  i.e. does a truly stable active
 phase exist?

\begin{figure}[t]
 \includegraphics[width=0.32\textwidth]{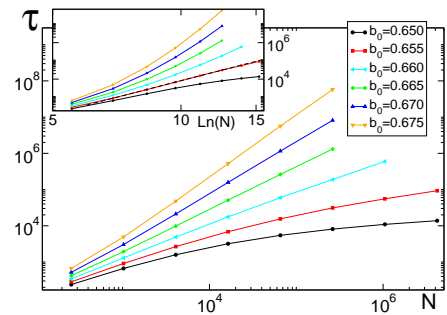}
 \caption{Main: Log-log plot of the lifetime $\tau$ as a function of system
   size $N$ for the TD-CP in $d=2$ for various $b_0$ and $\sigma=0.4$.
There is a finite region, $b_0 \in [0.656, 0.675]$ with generic
algebraic scaling of $\tau(N)$ and continuously varying
exponents. Inset: log-log plot of $\tau(N)$ vs $\ln(N)$; from the fit
at criticality (dashed line) we estimate $\tau \sim (\ln
N)^{5.18(5)}$.}
\label{CP2}
\end{figure}

 Let us first report on numerical simulations of the TD-CP performed
 in dimensions $d=1$, $d=2$, and $d \rightarrow \infty$ (for which we
 consider a fully connected network (FCN)). $b(t)$ is independently
 extracted at each Monte Carlo step from a homogeneous distribution $
 b\in [b_0 -\sigma, b_0+\sigma]$.  We performed both, homogeneous
 initial density experiments (with all sites initially active) and
 spreading ones (starting from a single active seed) \cite{AS}.  In
 the first set of experiments we measured the average value, over many
 realizations, of the density of activity $\rho(t)$ as a function of
 time and the mean lifetime $\tau(N)$ vs system size $N$.  In the
 second set, we measured standard quantities such as the survival
 probability as a function of time, $P_s(t)$ \cite{AS}.  Searching for
 power-laws of the form $\rho(t) \sim t^{-\theta}$ and $P_s(t) \sim
 t^{-\delta}$, we determined the critical point location $b_{0,c}=
 0.907(2)$, $0.656(1)$, and $0.500(1)$, and the exponent $\delta
 \approx 0.10(5) $, $0.126(2)$, and $0.5$, for dimensions $d=1$,
 $d=2$, and the FCN respectively (in all cases $\theta \approx
 \delta$; $\sigma=0.55$ for $d=1$ and $\sigma=0.4$ otherwise). For a
 fixed value of $\sigma$, the shift $b_{0,c}-b_{c,pure}$ is larger in
 $d=1$ than in $d=2$, and vanishes in $d=\infty$. Except for the
 mean-field value, and in agreement with previous findings
 \cite{Jensen}, these critical exponents are non universal, as they
 decrease upon increasing the noise amplitude $\sigma$.  Remarkably,
 in $d=2$ and $d=\infty$, $\tau(N)$ scales at criticality as $\tau
 \sim (\ln N)^{z'}$ with $z'(d=2)= 5.18(5)$ (inset of Fig.\ref{CP2})
 and $z'(d=\infty)= 3.49(5)$ for $\sigma=0.2$. The values of $z'(d)$
 do not seem universal either, as they decrease with increasing
 $\sigma$. Instead, in $d=1$ we observe standard power-law scaling
 $\tau \sim N^{1.55(1)}$.  Furthermore, in $d=2$ and $d=\infty$ (but,
 again, not in $d=1$), we find a whole region within the active phase
 ($b > b_{0,c}$) in which $\tau(N)$ grows generically as a power-law
 with continuously varying exponent $\zeta$, $\tau(N) \sim N^\zeta$,
 with $\zeta \rightarrow 0$ as $b_0 \rightarrow b_{0,c}^+$ (observe, in
 Fig.\ref{CP2}, the slight downward curvature in the $\log \tau - \log
 N$ curves at criticality, reflecting the asymptotic logarithmic
 behavior). Let us remark that obtaining data for larger sizes and
 deeper into the active phase, where the surviving times are huge,
 becomes excessively expensive. Hence, estimating with accuracy the
 upper limit of the algebraic scaling region is prohibitive. We have
 also measured $\tau(N)$ in the absorbing state; as in the pure model
 case, it scales as $\tau(N) \sim \ln(N)$ in all dimensions.
In summary, while the behavior of $\tau(N)$ in $d=1$ is similar to
that of pure systems, in $d=2$ and $d \rightarrow \infty$, we found
{\bf (i)} logarithmic scaling at criticality and {\bf (ii)} an
extended region with algebraic scaling.

\noindent
Let us now present analytical calculations for the high-dimensional limit
(FCN). Given that, at every single step, the change on the global
density $\rho$ is $\pm 1/N$, one can map its dynamics into a random
walk in the interval $[0,1]$, with jumps $\pm 1/N$ occurring with
probabilities $b(t) \,\rho (1-\rho)$ and $\left[1-b(t)\right] \rho$
respectively. The Master Equation for this process is easily written
\cite{Gardiner}, and by performing a $1/N$ expansion one readily
obtains a Fokker-Planck equation whose (Ito) Langevin equivalent is
(up to leading order),
\begin{equation}
\dot{\rho}(t)= a \rho - b\rho^2 + \alpha \sqrt{\rho}\, \eta(t) + \sigma 
\rho \, \xi(t),
\label{LangevinMF}
\end{equation}
with $a=2b_0-1+\sigma^2/2$, $b=b_0$, $\alpha=1/\sqrt{N}$, and the
noise $\xi(t) = 2(b(t)-b_0)/\sigma$.  Observe the presence of both, a
{\it demographic} noise, proportional to $\sqrt{\rho}$ which vanishes
in the $N \rightarrow \infty $ limit and an external or environmental
noise, linear in $\rho$ \cite{Hors,MN}. Generalizing to any spatial
dimension one can show that the corresponding Langevin equation is
just that of the directed percolation universality class \cite{AS} with
a fluctuating linear-term parameter:
\begin{equation}
\dot{\rho} = (a + \sigma \xi(t)) \rho - b\rho^2 + \nabla^2 \rho({\bf x},t) 
+ \gamma \sqrt{\rho} \, \eta({\bf x},t),
\label{Langevin}
\end{equation}
where $\gamma$ is a constant.  For the sake of generality, we have
numerically integrated Eq.~(\ref{Langevin}) in $d=1$, $d=2$ and in a
FCN, by using the scheme in \cite{Dornic}. We reproduced all the
findings above proving that our conclusions are robust, and apply to
any model in the directed percolation class not just the TD-CP.

In the case of uncorrelated white noise (as used in the numerics
above) the quasi-stationary solution of Eq.~(\ref{LangevinMF}) is $
P(\rho) \sim \rho^{-1} \left(1+\sigma^2 \rho/\alpha^2
\right)^{2\frac{b \alpha^2+a \sigma^2}{\sigma^4}-1} \exp(-2b
\rho/\sigma^2)$ which in the large $N$ limit can be approximated by
$P(\rho) \sim \rho^{2(a/\sigma^2 -1)} e^{-2b \rho/\sigma^2}$,
exhibiting an $a/\sigma^2$-dependent singularity at $\rho=0$. In this
limit, the singularity is not integrable for $a < a_c= \sigma^2/2$ ($b
< b_{0,c}(FCN)=0.5$ for all $\sigma$), and the only solution is a
delta distribution at $\rho=0$, i.e. the system is absorbing. Instead,
for finite $N$, there is a $1/\rho$ singularity for any value of $a$
and the only possible steady state is the absorbing one (as occurs for
any finite system with demographic noise \cite{AS}).
Defining $z=\ln \rho$, Eq.~(\ref{LangevinMF}) becomes
$\dot{z}=\tilde{a} - b \exp(z) + \sigma \xi(t)$, with
$\tilde{a}=a-\sigma^2/2$, which describes a random walker trapped in a
potential $V(z)=-\tilde{a} z + b \exp(z)$. It exhibits the three
following regimes: {\bf (i) Active phase ($\tilde{a}>0$)}: the time
required for the active state to fluctuate and reach the vicinity of
the absorbing state (which is approximately $\rho=1/N$, i.e. $z=-
\ln(N)$) and, eventually, die, is exponential in the height of the
potential \cite{Gardiner}, $\tau(N) \sim \exp \left[ V(-\ln
  N)/(\sigma^2/2) \right] \sim \exp (2 \tilde{a} \ln N /\sigma^2) \sim
N^{2 \tilde{a}/\sigma^2}$. This is, $\tau(N)$ exhibits generic
algebraic scaling with continuously varying exponents\cite{Leigh}.
Hence, the active phase is truly stable when $N \rightarrow \infty$.
{\bf (ii) Critical point ($\tilde{a}=0$)}: For sufficiently small
values of $z$ we have a free random walk (no potential barrier to be
overcome) which covers a typical distance $\sqrt{\tau}$ in time
$\tau$; equating this distance to $z=\ln N$, the time to die scales
logarithmically, $\tau \sim \left[ \ln(N)\right]^2$. {\bf (iii)
  Absorbing phase ($\tilde{a}<0$)}: $z$ decays linearly in time and,
hence, the time needed to reach $z=-\ln N$ scales as $\tau \sim \ln N
$.  These predictions are in excellent agreement with the
corresponding numerical results for the FCN.

Result {\bf (i)} can be recovered by using the path-integral
representation of Eq.~(\ref{LangevinMF}) \cite{AS}. The most probable
path to the absorbing state can be easily calculated in semiclassical
approximation. $\tau$ is simply the inverse of the probability weight
associated with such a path.  By using this formalism, Kamenev et
al. \cite{Kamenev} have recently investigated in an interesting work
the effect of correlated temporal disorder on a one-variable
birth-death process.  They conclude that, in the case of interest here
(short-time correlated noise), $\tau(N)$ grows exponentially with $N$
for weak noise amplitudes and algebraically in the strong
external-noise limit. Our result differs slightly from this: given
that the demographic noise amplitude in Eq.~(\ref{LangevinMF})
vanishes in the large $N$-limit, the strong-noise limit does not need
to be invoked to obtain algebraic scaling.


In order to extend our conclusions to finite dimensions and to make a
parallelism between the reported broad regions of generic algebraic
scaling \texttwelveudash that we call {\it temporal Griffiths phases}
(TGPs) \texttwelveudash and standard GPs, let us sketch the main
properties of GPs for the contact process equipped with quenched
disorder, i.e.  $b \rightarrow b({\bf x})$ \cite{QCP}. In the quenched
CP, rare regions with $b({\bf x}) > b_{c,pure}$ and arbitrary size $s$
appear with probability $\exp(-\alpha s)$, where $\alpha$ is a
disorder-dependent constant. Such regions are locally active and,
hence, activity survives on them until a coherent fluctuation kills
it. This occurs at a characteristic time $t_c(s) \sim \exp(\beta s)$
where $\beta$ is a constant, as given by the Arrhenius law
\cite{Gardiner}. Hence, the time-decay of the survival probability of
a homogeneous initial condition is given by the convolution $P_s(t)
\propto \int ds \exp(-\alpha s) \exp(-t/t_c(s)),$ and the leading
contribution in saddle point approximation comes from size
$s^*(t)=(1/\beta) \ln(\beta t / \alpha)$, implying $P_s(t) \propto
t^{-\alpha / \beta}$ (right at the critical point, the exponent
vanishes, and there is ``activated scaling'', characterized by a
logarithmic decay $P_s(t) \sim (\ln t)^{-\theta'}$
\cite{Vojta}). Similar expressions apply to the time decay of other
quantities such as the activity density, as well as to many different
systems with quenched disorder \cite{Vojta}.

\noindent
Thus, some analogies between GPs and TGPs are: {\it i}) In GPs
disorder is ``quenched in space''; in TGPs it is ``quenched in time''.
{\it ii}) In GPs rare (locally active) regions exist even if the
overall state is absorbing; in TGPs rare (temporarily absorbing)
time-intervals exist even if the overall state is active; i.e. the
roles of active/absorbing phases are exchanged.  {\it iii}) In GPs the
probability for a (rare) active region of size $s$ to occur is
$\exp(-\alpha s)$; in TGPs (rare) time intervals of length $T$ are
absorbing with probability $\exp(-\alpha T)$; hence, the typical time
to observe them is $\tau \sim \exp(\alpha T)$.  {\it iv}) In GPs, as
we just argued, the leading contribution of the decay at time $t$
comes from a rare region of size $s^* \sim \ln(t \beta/\alpha)/\beta$;
this combined with ({\it iii}) leads to a {\it generic power-law decay
  in time}, $t^{-\alpha/\beta}$.  In TGPs, the time required to reach
the absorbing state in an absorbing time-interval, is given by
$\exp(-\beta t^*) \sim 1/N$, or $t^* \sim \ln(N)/\beta$.  Equating
$t^*$ with $T$ in ({\it iii}), one obtains a {\it generic algebraic
  decay in system size}; $\tau(N) \sim N^{\alpha/\beta}$.  In
conclusion, TGPs are analogous to GPs by exchanging the roles of space
and time. These heuristic arguments seem to be valid even for finite
dimensional systems (down to $d=2$). The reason why a TGP phase is not
observed in $d=1$ is not completely clear to us. Presently we are
developing a semiclassical approximation, analogous to that in
\cite{Kamenev}, but for the spatialy explicit Eq.~(\ref{Langevin})
(see \cite{Fogedby}), in order to have a more precise understanding of
low-dimensional cases.

To further delve into the GP/TGP analogy, and given that GPs exhibit
generic divergences of the susceptibility \cite{Griffiths,GP}, we have
measured numerically the susceptibility, defined as $\Xi=\partial
\rho_{st}(h) / \partial h |_{h \rightarrow 0}$ where $\rho_{st}$ is
the average value of the activity in the steady state after
introducing an external field $h$ coupled to the system's dynamics,
for Eq.~(\ref{LangevinMF}) and for Eq.~(\ref{Langevin}) in $d=2$.
\begin{figure}[t]
\includegraphics[width=0.32\textwidth]{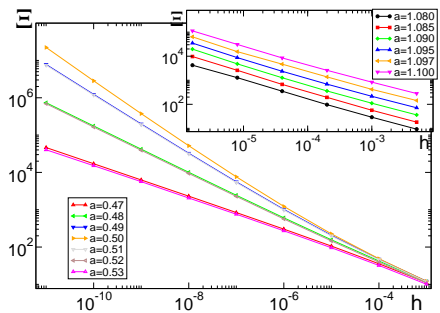}
\caption{Log-log plot of the susceptibility, $\Xi$, as a function of
  the field $h$, obtained by integrating Eq.~(\ref{LangevinMF}) (main
  plot) and Eq.~(\ref{Langevin}) for $d=2$ (inset), for different
  values of $a$.}
\label{SUS}
\end{figure}
Fig.\ref{SUS} shows that $\Xi$ measured from Eq.~(\ref{LangevinMF})
(which perfectly agrees with that in Monte Carlo simulations of the
TD-CP on FCN) diverges all along the TGP. Fig.\ref{SUS}-inset shows
generic divergences also for $d=2$; however, given that very small
values of $h$ cannot be reached in this case, it is difficult to
elucidate numerically whether such a divergence is for real in the $h
\rightarrow 0$ limit or it just a transient effect.

To analytically understand these findings, we take
Eq.~(\ref{LangevinMF}) in the $N \rightarrow \infty$ limit and include
an external field $h$. A new term $\exp\left[-2h/(\rho
  \sigma^2)\right]$,
 exhibiting an essential singularity at $\rho=0$, appears in
the stationary solution (see above). It is a matter of algebra to
verify analytically that $\partial \rho_{st}(h) / \partial h $
diverges algebraically as $h^{-1+2 |a-a_c|/\sigma^2}$, in an extended
interval $a \in [0,\sigma^2]$ around the critical point \cite{MN}.
If, in this calculation, we replace $\alpha$ by $\gamma$ which does
not vanish when $N \rightarrow \infty$ and mimics what happens in
finite dimensions (see Eq.(\ref{Langevin})), the parameter-dependent
singularities are replaced by the usual $\rho^{-1}$ absorbing state
singularity. This suggests that the generic divergence of the
susceptibility is a transient effect in the presence of non-vanishing
demographic noise. In that case, the strong external-noise limit needs
to be taken for the generic divergence to survive. Going beyond
mean-field, it can be proved by using simple field-theoretical
arguments (similarly to \cite{MN}) that Eq.~(\ref{Langevin}) with
$\alpha=0$ exhibits generic divergences of the susceptibility in a
broad interval even in finite spatial dimensions.
 
In summary, systems with absorbing states and fluctuating external
conditions exhibit a region in the active phase \texttwelveudash the
``temporal Griffiths phase''\texttwelveudash such that the mean
lifetime scales generically as a power-law (with continuously varying
exponents) of system size and logarithmically at criticality. This
occurs not only in mean field \cite{Leigh,Kamenev} but also in
extended systems as long as $d \geq 2$.  TGPs have deep analogies with
standard GPs, but the roles of space and time are reversed: in GPs
(TGPs) spatial (temporal) disorder leads to generic algebraic scaling
as a function of time (size). Moreover, as GPs, TGPs exhibit (at least
in the strong noise limit) generic divergences of magnitudes such as
the susceptibility and of stationary distribution functions.

TGPs could be measured in the experimental realizations of the
directed percolation class with liquid-crystals \cite{Kazz} by
introducing externally changing fields, and could appear in many other
systems such as in bistable Ising-like models with randomly changing
conditions. We hope this work will stimulate new research along these
lines.

We acknowledge the MICINN(FEDER) (FIS2007-60327 and FIS2009-08451),
DARPA grant HR0011-09-1-055 and J. de Andaluc{\'\i}a P09-FQM4682 for
support.


\vspace{-0.5cm}
\begin{thebibliography}{99}

\bibitem{Geoff} G. Grinstein and A. Luther, Phys. Rev.B{\bf 13},
  1329 (1976).


\bibitem{Griffiths} R. B. Griffiths, Phys. Rev. Lett. {\bf 23}, 17 (1969).

\bibitem{GP} A. J. Bray, Phys. Rev. Lett. {\bf 59}, 586 (1987).
 D.S. Fisher, Phys. Rev. Lett. {\bf 69}, 534 (1992).


\bibitem{Vojta} T. Vojta, J. Phys. A: Math. Gen. {\bf 39}, R143 (2006).



\bibitem{Odor} M. A. Mu\~noz {\it et al.},
 Phys. Rev. Lett. {\bf 105}, 128701  (2010).


\bibitem{SI} G.  Grinstein,
J. Appl. Phys.  {\bf 69}, 5441 (1991). 

\bibitem{foot} Spatio-temporal disorder might be needed in some cases.

\bibitem{Leigh} E. G. Leigh Jr. , J. Theor. Biol. {\bf 90}, 213
  (1981).



\bibitem{Hors}
J. Garc{\'\i}a-Ojalvo and J. M. Sancho, {\it Noise in Spatially Extended
  Systems} (Springer, N.Y. 1999).



\bibitem{Juanjo} J.J. Alonso, and M. A. Mu\~noz, 
EPL {\bf 56}, 485 (2001).

\bibitem{Jensen} I. Jensen, Phys. Rev. Lett. {\bf 77}, 4988 (1996).





\bibitem{Kamenev} A. Kamenev, B. Meerson, and B. Shklovskii,
  Phys. Rev. Lett. {\bf 101}, 268103 (2008). 

\bibitem{AS} H. Hinrichsen,
  Adv. Phys. {\bf 49}, 815 (2000). 



\bibitem{Gardiner}
C. W. Gardiner, {\it Handbook of Stochastic Methods}, Springer-Verlag,
Berlin and Heidelberg, 1985.





\bibitem{MN} G. Grinstein, M.A. Mu{\~{n}}oz and Y. Tu,
  Phys. Rev. Lett.  {\bf 76}, 4376 (1996).  




\bibitem{Dornic} 
I. Dornic, H. Chat\'e, and  M. A. Mu\~noz,
Phys. Rev. Lett. {\bf 94}, 100601 (2005). 



\bibitem{QCP} A. J. Noest, Phys. Rev. Lett. {\bf 57}, 90 (1986).


\bibitem{Fogedby} H. C. Fogedby, J. Hertz, and A. Svane,
  EPL {\bf 62},  795 (2003). 

\bibitem{Kazz} K. A. Takeuchi, {\it et al.}
  Phys. Rev. Lett. 99, 234503 (2007).


\end{thebibliography}
\end{document}